\documentclass[twocolumn,twocolappendix]{aastex7}
\usepackage{url}
\usepackage{xspace}
\usepackage{lipsum}

\let\oldAA\AA
\renewcommand{\AA}{{\text{\normalfont\oldAA}}}
\newcommand{\eg}{\textit{e.g.}}

\newcommand{\target}{CR3\xspace}
\usepackage{xcolor}
\usepackage{booktabs}
\newcommand{\lya}{Ly$\alpha$}
\newcommand{\ha}{H$\alpha$}
\newcommand{\hb}{H$\beta$}
\newcommand{\oiii}{[\ion{O}{3}]}
\newcommand{\heii}{\ion{He}{2}}
\newcommand{\hei}{\ion{He}{1}}
\newcommand{\civ}{\ion{C}{4}}

\shorttitle{}
\shortauthors{}
\graphicspath{{./}{figures/}}
\begin{document}

\title{
A Metal-Free Galaxy at $z = 3.19$? Evidence of Late Population III Star Formation at Cosmic Noon
}

\author[0009-0003-4133-0292]{Sijia Cai}
\affiliation{Department of Astronomy, Tsinghua University, Beijing 100084, China}
\email{caisj23@mails.tsinghua.edu.cn}

\author[0000-0001-6251-649X]{Mingyu Li}
\affiliation{Department of Astronomy, Tsinghua University, Beijing 100084, China}
\email{lmy22@mails.tsinghua.edu.cn}

\author[0000-0001-8467-6478]{Zheng Cai}
\affiliation{Department of Astronomy, Tsinghua University, Beijing 100084, China}
\email[show]{zcai@tsinghua.edu.cn}

\author[0000-0003-0111-8249]{Yunjing Wu}
\affiliation{Department of Astronomy, Tsinghua University, Beijing 100084, China}
\email{yunjingwu@arizona.edu}

\author[0000-0002-3489-6381]{Fujiang Yu}
\affiliation{Department of Astronomy, Tsinghua University, Beijing 100084, China}
\email{yufj@mail.tsinghua.edu.cn}

\author[0000-0001-5414-5131]{Mark Dickinson}
\affiliation{NSF’s National Optical-Infrared Astronomy Research Laboratory, 950 N. Cherry Ave., Tucson, AZ 85719, USA}
\email{mark.dickinson@noirlab.edu}

\author[0000-0002-4622-6617]{Fengwu Sun}
\email{fengwu.sun@cfa.harvard.edu}
\affiliation{Center for Astrophysics $|$ Harvard \& Smithsonian, 60 Garden St., Cambridge, MA 02138, USA}

\author[0000-0003-3310-0131]{Xiaohui Fan}
\affiliation{Steward Observatory, University of Arizona, 933 N Cherry Avenue, Tucson, AZ 85721, USA}
\email{xiaohuidominicfan@gmail.com}

\author[0000-0003-4877-1659]{Ben Wang}
\affiliation{Department of Astronomy, Tsinghua University, Beijing 100084, China}
\affiliation{Leiden Observatory, Leiden University, Leiden 2333 CA, Netherland}
\email{wb20@mails.tsinghua.edu.cn}

\author[0000-0002-3736-476X]{Fergus Cullen}
\affiliation{Institute for Astronomy, University of Edinburgh, Royal Observatory, Edinburgh EH9 3HJ, UK}
\email{fergus.cullen@ed.ac.uk}

\author[0000-0002-1620-0897]{Fuyan Bian}
\affiliation{European Southern Observatory, Alonso de Córdova 3107, Casilla 19001, Vitacura, Santiago 19, Chile}
\email{fuyan.bian@eso.org}

\author[0000-0001-6052-4234]{Xiaojing Lin}
\affiliation{Department of Astronomy, Tsinghua University, Beijing 100084, China}
\email{linxj21@mails.tsinghua.edu.cn}

\author[0000-0001-9189-0368]{Jiaqi Zou}
\affiliation{Department of Astronomy, Tsinghua University, Beijing 100084, China}
\email{zoujq20@mails.tsinghua.edu.cn}

\correspondingauthor{\href{mailto:zcai@tsinghua.edu.cn}{Zheng Cai}}

% \collaboration{all}{The Terra Mater collaboration}

%% Use the \collaboration command to identify collaborations. This command
%% takes an optional argument that is either a number or the word "all"
%% which tells the compiler how many of the authors above the command to
%% show. For example "\collaboration[all]{(DELVE Collaboration)}" wil include
%% all the authors above this command.
%%
%% Mark off the abstract in the ``abstract'' environment. 
\begin{abstract}

Star formation from metal-free gas, the hallmark of 
the first generation of Population III stars, was long 
assumed to occur only in the very early Universe. We report 
the discovery of MPG-CR3 (Metal-Pristine 
Galaxy COSMOS Redshift 3; hereafter CR3), an extremely metal-poor galaxy 
at redshift $ z= 3.193\pm0.016$.
From JWST, VLT, and Subaru
observations, 
CR3 exhibits exceptionally strong Ly$\alpha$, 
H$\alpha$, and \hei$\,\lambda$10830 emission. 
We measure rest-frame equivalent widths of 
EW$_0$(Ly$\alpha$) $= 822\pm101\AA$ and EW$_0$(H$\alpha$) $= 2814\pm327\AA$, 
among the highest seen in star-forming systems. 
No metal lines, e.g. \oiii$\,\lambda\lambda4959,5007$, \civ$\,\lambda\lambda1548,1550$,  
have statistically significant detections, 
placing a 2-$\sigma$ upper limit on the gas-phase metallicity of $12+\log(\mathrm{O/H}) < 6.52$ ($Z < 7 \times 10^{-3}\,Z_{\odot}$) with strong-line calibration established by JWST, making it the most metal-poor galaxy known at cosmic noon. Considering systematic uncertainties of $\gtrsim 0.3$ dex in the calibrations, the most conservative 2-$\sigma$ upper limit is set to $12+\log(\mathrm{O/H}) < 6.95$.
The observed Ly$\alpha$/H$\alpha$ flux ratio is $13.9\pm2.5$, 
indicating negligible dust attenuation. 
Spectral energy distribution modeling with Pop III stellar templates indicates 
a very young ($\sim2$ Myr), low-mass ($M_* \approx 6.1\times 10^5 M_\odot$) stellar population. Further, 
the photometric redshifts reveal that 
\target could reside in a slightly underdense 
environment ($\delta \approx -0.12$).  
\target provides evidence that first-generation star formation 
could persist well after the epoch of reionization,  
challenging the 
conventional view that 
pristine star formation ended by $z\gtrsim6$. 
\end{abstract}

%% Keywords should appear after the \end{abstract} command. 
%% The AAS Journals now uses Unified Astronomy Thesaurus (UAT) concepts:
%% https://astrothesaurus.org
%% You will be asked to selected these concepts during the submission process
%% but this old "keyword" functionality is maintained in case authors want
%% to include these concepts in their preprints.
%%
%% You can use the \uat command to link your UAT concepts back its source.
% \keywords{\uat{Interacting galaxies}{802} --- \uat{Jets}{870} --- \uat{Interstellar medium}{847} --- \uat{Starburst galaxies}{1570}}

%% From the front matter, we move on to the body of the paper.
%% Sections are demarcated by \section and \subsection, respectively.
%% Observe the use of the LaTeX \label
%% command after the \subsection to give a symbolic KEY to the
%% subsection for cross-referencing in a \ref command.
%% You can use LaTeX's \ref and \label commands to keep track of
%% cross-references to sections, equations, tables, and figures.
%% That way, if you change the order of any elements, LaTeX will
%% automatically renumber them.

\section{Introduction} \label{sec:intro}

Population III (Pop III) galaxies are theoretically predicted to be the first generation of stellar systems formed from metal–free primordial gas shortly after the Big Bang \citep{Klessen2023ARA&A..61...65K}. In the absence of elements heavier than helium, the interstellar medium (ISM) is expected to lack metal–line cooling, leading in models to higher gas temperatures and preferential formation of very massive (\(\gtrsim100\,M_\odot\)), hot (\(T>10^5\) K) stars (\citealt{Tumlinson2000ApJ...528L..65T}; \citealt{Schaerer2003A&A...397..527S}). 
These zero–metallicity stars are predicted to produce extremely hard ionizing spectra, with an excess of photons above the He\(^+\) ionization threshold (E $>$ 54.4 eV), which would in turn drive exceptionally strong hydrogen recombination lines (\lya, \ha, \hb) and nebular \heii\ emission (e.g. \heii\(\,\lambda1640\) with rest–frame equivalent width \(\gtrsim5\) \AA), while the absence of heavy elements would imply no detectable metal lines such as \oiii~or \ion{C}{3}] \citep{Bromm2011ARA&A..49..373B}. 
Over the past two decades, the quest to identify Pop III galaxy candidates has driven significant advances in both instrumentation and methodology, with the promise that confirmed detections would tightly constrain the primordial initial mass functions (IMFs), the timeline of cosmic reionization, and the onset of chemical enrichment in the intergalactic medium.  

Early attempts using ground‐based telescopes focused on deep narrow‐band imaging and spectroscopy to isolate unusually strong \heii\(\,\lambda1640\) emission \citep[\eg,][]{Cassata2013A&A...556A..68C,Sobral2015ApJ...808..139S,Nanayakkara2019A&A...624A..89N,Saxena2020A&A...636A..47S,Gonz'alez-D'iaz2025arXiv250611685G}.
However, nebular \heii~can also arise in harder radiation fields produced by active galactic nuclei or Wolf–Rayet (WR) star populations (\citealt{Leitherer1999ApJS..123....3L}; \citealt{Crowther2007ARA&A..45..177C}); thus, robust Pop III identification requires multi‐line diagnostics (e.g.\ absence of high‐ionization metal lines, line‐width constraints, and diagnostic line‐ratio diagrams) to exclude these alternative ionizing sources.
Studies of nearby metal-poor galaxies have revealed narrow nebular \heii\ emission in systems lacking obvious AGN or WR signatures, suggesting that very hard ionizing spectra can arise in local environments \citep[\eg,][]{Shirazi2012MNRAS.421.1043S,Kehrig2015ApJ...801L..28K,Senchyna2017MNRAS.472.2608S}.
Since the deployment of space‐based near‐infrared spectrographs such as HST/WFC3 and JWST/NIRSpec, sensitivity to \heii\(\,\lambda1640\) emission at \(z\gtrsim6\) has improved dramatically \citep[\eg,][]{Cai2011ApJ...736L..28C,Cai2015ApJ...799L..19C,Wang2024ApJ...967L..42W,Maiolino2024A&A...687A..67M}. 

It is widely expected that truly metal‐poor galactic environments become more prevalent at \(z\gtrsim6\) 
\citep{Xu2016ApJ...823..140X,Sarmento2018ApJ...854...75S,Jaacks2019MNRAS.488.2202J,Liu2020MNRAS.497.2839L,Skinner2020MNRAS.492.4386S,Visbal2020ApJ...897...95V,Sarmento2022ApJ...935..174S,Venditti2023MNRAS.522.3809V,Zier2025arXiv250303806Z}, and the advent of JWST has intensified searches in this early epoch.  
For example, NIRSpec integral‐field spectroscopy (IFS) revealed the strongly lensed Pop III candidate stellar complex ``LAP1" amplified by the Hubble Frontier Field galaxy cluster MACS J0416 at  \(z=6.6\) \citep{Vanzella2023A&A...678A.173V}. 
Its most metal‐deficient clump, ``LAP1–B", was further observed with NIRSpec medium‐resolution gratings to have a gas‐phase metallicity as low as \(\sim4.2\times10^{-3}\,Z_\odot\) (\citealt{Nakajima2025arXiv250611846N}).  
In parallel, \citet{Fujimoto2025arXiv250111678F} developed a novel photometric selection for Pop III candidates at \(z\sim6\text{--}7\) by combining strong \ha~emission, a pronounced Balmer jump, and anomalously weak \oiii~lines.  
Meanwhile, NIRCam wide-field slitless spectroscopy has demonstrated that deep grism data can efficiently select extremely metal-poor galaxies or clumps via the \oiii$\,\lambda$5007/\hb~ratio \citep[e.g.,][]{Naidu2024arXiv241001874N, Hsiao2025arXiv250503873H,Morishita:2025zvd}, albeit high-fidelity metallicity measurements will rely on further deep NIRSpec confirmation.
To date, perhaps owing to their intrinsically low stellar masses, extremely faint continua, and the expected short-lived nature of the Pop III phase, Pop III host candidates remain rare even in the JWST era.

From theoretical perspectives, cosmological simulations 
suggest that Pop III star formation can extend well beyond reionization into much lower redshifts. 
\citet{Tornatore2007MNRAS.382..945T} showed that inefficient metal mixing and slow enrichment during hierarchical growth allow Pop III star formation to persist down to \(z\sim2.5\), albeit at a low rate of \(\sim10^{-5}-10^{-7}\,M_\odot\,\mathrm{yr}^{-1}\,\mathrm{Mpc}^{-3}\). At these redshifts, such star formation is expected to occur in the low‐density outskirts of halos, where long free‐fall timescales suppress efficient star formation. 
Indeed, absorption‐line studies of damped Ly$\alpha$ systems (DLAs) at \(z\sim2\)--3 have uncovered extremely metal‐poor gas—exhibiting carbon enhancement and \(\alpha\)‐element suppression indicative of Pop III enrichment—and even traced reservoirs as pristine as \([\mathrm{Fe}/\mathrm{H}]\approx-3.5\) \citep{Erni2006A&A...451...19E,Cooke2011MNRAS.412.1047C,Cooke2011MNRAS.417.1534C,Kulkarni2013ApJ...772...93K,Fumagalli2015MNRAS.446.3178F,Cooke2017MNRAS.467..802C,Welsh2023MNRAS.525..527W}.
More recently, \citet{Mondal2025arXiv250606831M} identified the \heii\,\(\lambda1640\) emitter GNHeII J1236+6215 at \(z=2.98\), suggesting that its signal could be driven by Pop~III stars during the cosmic noon epoch, while \citet{Vanzella2025arXiv250907073V} reported the discovery of the pristine star-forming complex LAP2 at $z=4.19$.
Together, these results suggest that the \(z\sim2\text{--}5\) window may still harbor Pop III–dominated galaxies.  

In this study, we focus on a faint galaxy during the cosmic noon era that exhibits strong \ha~emission yet unusually faint \oiii$\,\lambda\lambda$4959,5007, and for which deep rest‐frame ultraviolet data (including \lya~and \heii$\,\lambda$1640) are available.
Here, we present the discovery of a Pop III galaxy candidate at \(z=3.193\) in the COSMOS field, which we designate MPG-CR3 (Metal-Pristine Galaxy COSMOS Redshift 3, hereafter \target\footnote{The acronym ``CR3" also coincides with the standard suffix “.CR3” used for raw image files, aligning with the interpretation that the galaxy’s chemical abundance is ``raw" (i.e., chemically unprocessed).}).  

The paper is structured as follows: Section \ref{sec:obs} describes the imaging and spectroscopic observations from JWST, Subaru, and VLT; Section \ref{sec:analysis} presents measurements of the emission-line properties, gas-phase metallicity, SED modeling, and the characterization of the large-scale environment of the \target. We discuss this Pop~III candidate galaxy at $z\sim3$ in Section \ref{sec:discuss}, and we summarize in Section \ref{sec:conclusion}.
Throughout this paper we assume a flat \(\Lambda\)CDM cosmology with parameters taken from \citet{Planck2020A&A...641A...6P}: \(H_0 = 67.7\)\,km\,s\(^{-1}\)\,Mpc\(^{-1}\), \(\Omega_{\rm m} = 0.31\), and \(\Omega_\Lambda = 0.69\).
In this cosmology, 1\arcsec\ corresponds to 7.73 kpc of physical length at $z=3.193$.
All magnitudes are given in the AB system \citep{Oke1983ApJ...266..713O}.
We adopt a solar oxygen abundance of $12 + \log(\mathrm{O/H}) = 6.89$ from \citet{Asplund2021A&A...653A.141A}.

\section{Observations} \label{sec:obs}

\begin{figure*}[htbp]
\centering
\includegraphics[width=\textwidth]{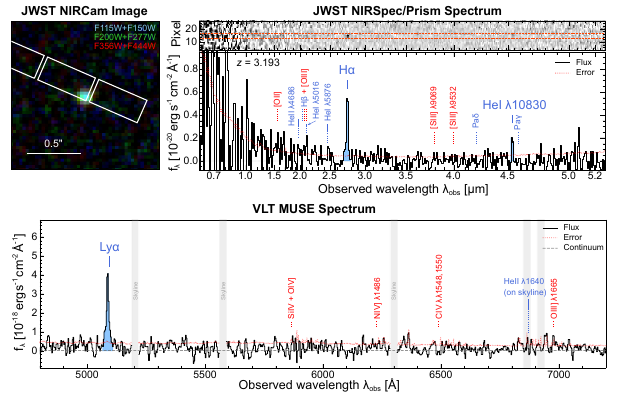} 
\caption{\textbf{JWST and VLT observation of the \target.} 
\textbf{Upper left:} Pseudo-color composite image constructed from JWST/NIRCam images, using blue (F115W, F150W), green (F200W, F277W), and red (F356W, F444W) channels. The orientation of the MSA shutters is indicated by the white region.
\textbf{Upper right:} JWST/NIRSpec prism 2D and 1D spectra. The red dashed lines indicate the optimal extraction aperture on the 2D spectrum. In the 1D panel, the black solid line shows the flux, while the red dotted line shows the associated error. The \ha\ and \hei$\,\lambda10830$ emission features are highlighted with blue shaded regions.  
\textbf{Lower:} 1D VLT/MUSE spectrum extracted from a circular aperture, rebinned and Hanning-smoothed. The black solid line shows the flux, and the red dotted line indicates the error. The Subaru/HSC $r$-band continuum level is overplotted as a gray dashed line. The \lya\ emission line is highlighted with a blue shaded region. Strong sky emission lines are marked with light gray shaded areas. Hydrogen and helium emission lines are labeled in blue, while metal lines are labeled in red.
}
\label{fig:spec}
\end{figure*}

\begin{figure*}[htbp]
\centering
\includegraphics[width=\textwidth]{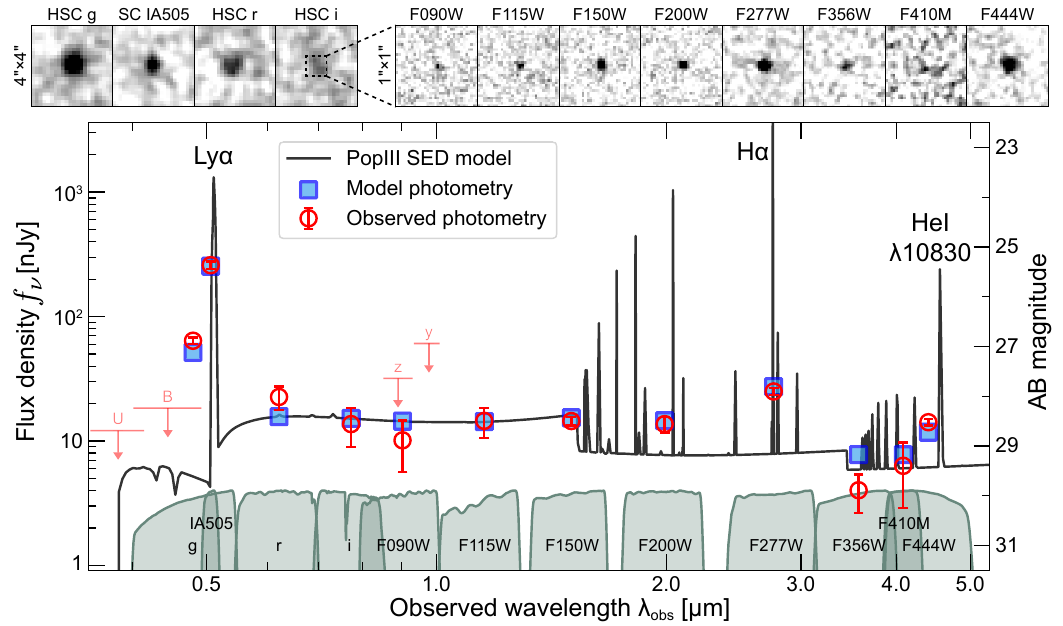}
\caption{\textbf{Multi-wavelength images and SED modeling of the \target.}  
\textbf{Upper:} Cutout stamps of the \target from Subaru/Suprime IA505, Subaru/HSC $g$, $r$, $i$, and JWST/NIRCam F090W, F115W, F150W, F200W, F277W, F356W, F410M, and F444W. The cutout image size is $4^{\prime\prime}$ for Subaru and $1^{\prime\prime}$ for JWST.
\textbf{Lower:} Spectral energy distribution (SED) modeling of the \target. The best-fit Pop~III SED model is shown as a black curve, with model photometry (blue squares) and observed photometry (blue circles) overlaid. The 2$\sigma$ upper limits in the U, B, $z$, and $y$ bands are indicated by red arrows. The Ly$\alpha$ emission boosts the fluxes in the IA505 and HSC-$g$ bands; H$\alpha$ emission boosts the F277W band; and \ion{He}{1} emission enhances the F444W band.
}
\label{fig:img+sed}
\end{figure*}

The \target is located in the COSMOS-CANDELS field \citep{Koekemoer2011ApJS..197...36K,Grogin2011ApJS..197...35G}, which is the core region of the COSMOS deep extragalactic field \citep{Scoville2007ApJS..172....1S}, with extensive archival observations and well-studied source catalogues.
The equatorial coordinates of \target are RA = 150.148631 deg and Dec = 2.435939 deg.
In this work, we utilize imaging and spectroscopic data from JWST,
Subaru Telescope, and Very Large Telescope (VLT), as detailed below and illustrated in Figure \ref{fig:spec} \& \ref{fig:img+sed}.

\subsection{Imaging Data}

\textbf{JWST imaging}. The \target is observed by four JWST imaging programs: COSMOS-Web \citep[GO\#1727; PIs: J. Kartaltepe and C. Casey;][]{Casey2023ApJ...954...31C}, PRIMER (GO\#1837; PI: J. Dunlop), COSMOS-3D (GO\#5893; PI: K. Kakiichi), and DDT\#6585 (PI: D. Coulter) using the Near-Infrared Camera \citep[NIRCam;][]{Rieke2023PASP..135b8001R} with filter bands of F090W, F115W, F150W, F200W, F277W, F356W, F410M, and F444W.
We reduce and coadd JWST imaging data from these four programs using the standard JWST pipeline\footnote{\url{https://github.com/spacetelescope/jwst}} version v1.17.1 \citep{bushouse_2025_14597407} with recommended reference file of \texttt{jwst\_1322.pmap}.
We perform an astrometric correction for each individual exposure to match the COSMOS2025 catalog \citep{Shuntov2025arXiv250603243S}, which has been registered to \textit{Gaia} data \citep{GaiaCollaboration2023A&A...674A...1G}.
The final mosaic images are drizzled with \texttt{pix\_frac=1.0} and a pixel size of 0\farcs03.
We note that the \target is also covered by JWST NIRISS pure parallel program GO\#4681 (PIs: K. Glazebrook and G. Brammer) with 3779.3s NIRISS imaging and 6957.426s NIRISS WFSS in F200W band, which we do not involve in this work.

\textbf{Subaru imaging}.
We use optical imaging data from the Hyper Suprime-Cam Subaru Strategic Program \citep[HSC-SSP,][]{Aihara2022PASJ...74..247A} and the Subaru COSMOS 20 project \citep{Taniguchi2007ApJS..172....9T,Capak2007ApJS..172...99C,Taniguchi2015PASJ...67..104T} with Suprime-Cam.
For the HSC-SSP data, we use the third public data release (PDR3) including g,r,i,z,y bands, which are downloaded from the PDR3 data access tool\footnote{\url{https://hsc-release.mtk.nao.ac.jp/doc/index.php/data-access__pdr3/}}.
Additionally, we incorporate data from Suprime-Cam, which are downloaded from the COSMOS Archive\footnote{\url{https://irsa.ipac.caltech.edu/data/COSMOS/overview.html}}.

\subsection{Spectroscopic Data}

\textbf{JWST NIRSpec data}.
The NIRSpec/MSA low-resolution prism spectra of the \target (source ID: 177138) were observed on May 22, 2025 by the CANDELS-Area Prism Epoch of Reionization Survey (CAPERS) program (GO\#6368; PI: M. Dickinson).
CAPERS is a JWST Cycle 4 Treasury program of deep NIRSpec PRISM spectroscopy in three CANDELS fields including COSMOS, which will build a spectroscopic legacy data set for as many as 10,000 galaxies up to $z>10$.
The total effective exposure time is 4.74 h.
We reduced the NIRSpec data using the STScI JWST pipeline (software version 1.17.1 and \texttt{jwst\_1322.pmap}) and the \texttt{msaexp}\footnote{\url{https://github.com/gbrammer/msaexp}} package \citep{Brammer2023zndo...8319596B}. We processed the raw NIRSpec data using the Level 1 steps of the standard STScI JWST pipeline. Subsequently, after applying the 1/f noise correction and identifying snowball artifacts, we performed the Level 2 processing using the \texttt{msaexp}. Background subtraction was achieved using the standard three-shutter nod offset strategy, yielding 2D spectra. Finally, we optimally extracted the 1D spectra from the 2D spectra. We determine the spectroscopic redshift to be $z=3.193\pm0.016$ by fitting the observed H$\alpha$ and \hei$\,\lambda$10830 emission lines with the emission-line templates provided in the \texttt{msaexp.spectrum} module. The redshift uncertainty is estimated from the shape of the $\chi^2$ distribution around the best-fit solution, corresponding to a 68\% confidence interval.

\textbf{VLT MUSE data}. We found that the \target is also covered by archival observations from the VLT/MUSE instrument. The data were obtained from the program ``MUSE's View of an Intense Star-forming Proto-cluster Candidate at $z \sim 4.2$'' (Program ID: 0102.A-0222; PI: K. Caputi). We utilized the publicly available, pipeline-processed datacube provided by the MUSE consortium\footnote{\url{https://archive.eso.org/dataset/ADP.2019-04-02T09:19:54.677}}. The observations were carried out on 10 March 2019, with an effective exposure time of 2659\,s and a median seeing of $0\farcs937$.

\section{Methods and Analysis} \label{sec:analysis}

\subsection{Emission Lines} \label{sec:emission-lines}

In the \target, Ly$\alpha$ emission is clearly detected. As shown in Figure~\ref{fig:img+sed}, both the Suprime-Cam IA505 intermediate band and HSC $g$-band photometry are significantly enhanced due to strong Ly$\alpha$ emission, which is independently confirmed by VLT/MUSE spectroscopy (Figure~\ref{fig:spec}).
To quantify the Ly$\alpha$ flux, we construct an emission line surface brightness (SB) map by integrating the MUSE datacube over the wavelength range corresponding to a $\pm2\sigma$ window of the best-fit Gaussian profile. As the spatial resolution of MUSE is limited by atmospheric seeing, the Ly$\alpha$ emission appears compact. We therefore adopt a circular aperture with radius 1$^{\prime\prime}$ (approximately twice the seeing FWHM) to capture the total flux while limiting background noise. The extracted 1D spectrum is smoothed with a Hanning kernel and rebinned with a step of 3. The continuum flux at Ly$\alpha$ wavelength is estimated from the HSC $r$-band photometry, which samples the rest-UV redward of Ly$\alpha$. 
A Gaussian fit to the Ly$\alpha$ emission yields a total flux of \( (5.8 \pm 0.7) \times 10^{-17}\,\mathrm{erg\,s^{-1}\,cm^{-2}} \) and a rest-frame equivalent width (EW$_0$) of \( 822 \pm 101\,\text{\AA} \), as summarized in Table \ref{tab:line}. At the wavelength of \heii$\,\lambda1640$, no emission is detected, as this feature coincides with an OH sky line. We therefore derive a $1\sigma$ upper limit on the flux of \(1.4 \times 10^{-17}\,\mathrm{erg\,s^{-1}\,cm^{-2}}\) and a corresponding EW$_0$ $\lesssim$ 180 \text{\AA}.

In addition, strong H$\alpha$ and \hei$\,\lambda10830$ emission lines are clearly evident in the NIRSpec prism spectrum, as well as in the observed excess flux in the NIRCam F277W and F444W bands (Figure~\ref{fig:spec} \& \ref{fig:img+sed}). We model the underlying continuum using a power-law fit and simultaneously fit the emission lines with single Gaussians. The inferred H$\alpha$ flux is \(f({\rm H\alpha}) = (1.8 \pm 0.2) \times 10^{-18}\,\mathrm{erg\,s^{-1}\,cm^{-2}}\), with \(\mathrm{EW}_0({\rm H}\alpha) = 2814 \pm 327\,\text{\AA}\), and the measured \hei$\,\lambda10830$ flux is $f$(\hei$\,\lambda10830$) \( = (4.5 \pm 1.5) \times 10^{-19}\,\mathrm{erg\,s^{-1}\,cm^{-2}}\), with EW$_0$(\hei$\,\lambda10830$) \(= 2360 \pm 807\,\text{\AA}\).

Given that the source falls near the edge of the MSA shutter, the H$\alpha$ line flux derived from the NIRSpec prism spectrum may be underestimated due to slit losses. We correct for this by convolving the spectrum with the NIRCam F277W filter curve and scaling it to match the F277W photometry, yielding a flux-corrected H$\alpha$ value of \( (4.2 \pm 0.6) \times 10^{-18}\,\mathrm{erg\,s^{-1}\,cm^{-2}} \). This leads to a Ly$\alpha$/H$\alpha$ ratio of \( 13.9 \pm 2.5 \).

\begin{table}%[htpb]
\centering
\caption{Emission Line Properties}
\label{tab:line}
\begin{tabular*}{\linewidth}{@{\extracolsep{\fill}}lcc}
\toprule
\toprule
Parameter & Value &  Unit \\
\midrule
\midrule
$f$(Ly$\alpha$) & $(5.8 \pm 0.7) \times 10^{-17}$ & \text{cgs}\footnote{Flux in \text{cgs} units: erg s$^{-1}$ cm$^{-2}$.} \\
$f(\rm{H}\alpha)$ & $(1.8 \pm 0.2) \times 10^{-18}$ & \text{cgs} \\
$f_{\rm{scaled}}$(H$\alpha$) & $(4.2 \pm 0.6) \times 10^{-18}$ & \text{cgs} \\
$f$(H$\beta$)\footnote{The \hb~measurement is based on Case B assumption.} & $(6.3 \pm 0.7) \times 10^{-19}$ & \text{cgs} \\
$f$(\hei$\,\lambda10830$) & $(4.5 \pm 1.5) \times 10^{-19}$ & \text{cgs} \\
$f$(\oiii$\,\lambda5007$) & $< 2.8 (5.6) \times 10^{-19}$ & \text{cgs}, 1(2)$\sigma$ \\
\midrule
EW$_0$(Ly$\alpha$) & $822 \pm 101$ & \AA \\
EW$_0$(H$\alpha$) & $2814 \pm 327$ & \AA \\
EW$_0$(\hei$\,\lambda10830$) & $2360 \pm 807$ & \AA \\
\midrule
R3 & $< 0.45(0.89)$  & 1(2)$\sigma$ \\
12 + log(O/H)\footnote{The oxygen metallicity is derived from the calibration by \citet{Sanders2024ApJ...962...24S}.} & $< 6.30(6.52)$  & 1(2)$\sigma$ \\
Ly$\alpha$/H$\alpha$ & $13.9 \pm 2.5$ & -- \\
\bottomrule
\bottomrule
\end{tabular*}
\end{table}

\subsection{Gas-phase Metallicity}

In the rest–UV to optical spectra of the \target we detect only \lya, \ha~and \hei$\,\lambda10830$ emission; unfortunately, the key Pop III tracer \heii$\,\lambda1640$ coincides with a strong OH sky line, precluding any reliable measurement of its flux.  No metal lines, such as %forbidden 
carbon, nitrogen, or oxygen lines are currently seen.

\begin{figure*}
    \centering
    \includegraphics[width=\textwidth]{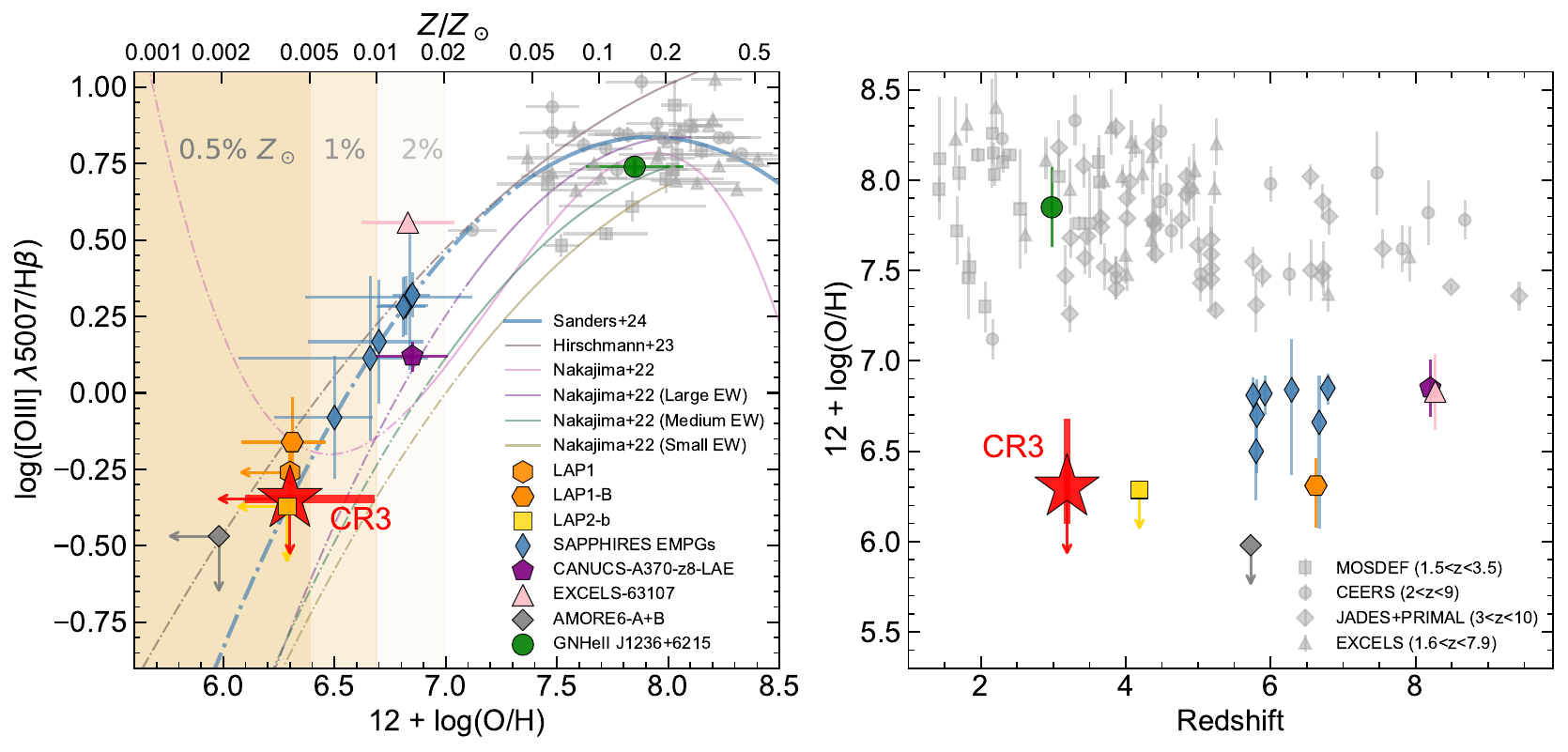} 
    \caption{\textbf{Constraints on the gas-phase metallicity of the \target.} \textbf{Left:} We apply the \citet{Sanders2024ApJ...962...24S} strong-line calibration to the measured upper limit on the R3 ratio (\oiii$\,\lambda5007$/H$\beta$; blue curve). For comparison, metallicity calibration curves from \citet{Hirschmann2023MNRAS.526.3504H} and \citet{Nakajima2022ApJS..262....3N} are presented. The dash-dotted extensions indicate the extrapolated regimes of the calibrations. Our measurement is compared to other known spectroscopically confirmed Pop~III candidates or extremely metal-poor galaxies (EMPGs), including the LAP1 stellar complex (orange hexagon, \citealt{Vanzella2023A&A...678A.173V}), the LAP1-B clump (orange hexagon, \citealt{Nakajima2025arXiv250611846N}), the LAP2-b stellar complex (yellow square, \citealt{Vanzella2025arXiv250907073V}), the AMORE6-A+B (gray diamond, \citealt{Morishita:2025zvd}), the GNHeII J1236+6215 (green circle, \citealt{Mondal2025arXiv250606831M}), the CANUCS-A370-z8-LAE (purple pentagon, \citealt{Willott2025arXiv250207733W}), the EXCELS-63107 (pink triangle, \citealt{Cullen2025MNRAS.540.2176C}) and the SAPPHIRES extremely metal-poor galaxies (blue diamond, \citealt{Hsiao2025arXiv250503873H}). Light gray markers show samples from the MOSDEF \citep{Sanders2020MNRAS.491.1427S}, the CEERS \citep{Sanders2024ApJ...962...24S}, the JADES and the PRIMAL \citep{Chakraborty2025ApJ...985...24C}, and the EXCELS \citep{Scholte2025MNRAS.540.1800S}. \textbf{Right:} Redshift–metallicity distribution of the sample, highlighting the extremely low metallicity of the \target compared to other galaxies at similar and higher redshifts. All limits shown represent 1-$\sigma$ upper bounds. The red rectangle indicates the systematic uncertainty associated with the strong-line calibration discussed in Section \ref{sec:discuss-4.1}.
}
    \label{fig:R3_Z}
\end{figure*}

In Figure \ref{fig:R3_Z}, we constrain the gas-phase oxygen abundance using the strong-line ratio R3 = $f$(\oiii$\,\lambda$5007)/$f$(\hb), which is the only available diagnostic in our data. Other commonly used diagnostics such as R23 = $f$([\ion{O}{2}])+$f$(\oiii)/$f$(\hb) are not constraining here due to the low signal-to-noise ratio (SNR) at the wavelength of [\ion{O}{2}]$\,\lambda3727$. At the low spectral resolution of the NIRSpec/prism data, H$\alpha$ may be blended with the neighboring [\ion{N}{2}] lines. However, at such low metallicity, [\ion{N}{2}]/H$\alpha$ is typically negligible ($\lesssim$ 0.01), and thus the H$\alpha$ flux is not significantly contaminated \citep[\eg,][]{Izotov1997ApJS..108....1I}.
\hb~is undetected in the low-SNR blue end of the prism spectrum. Section \ref{sec:emission-lines} (also discussed in Section \ref{sec:discuss-4.1}) shows that the observed Ly$\alpha$/H$\alpha$ ratio of $13.9\pm2.5$ demonstrates that dust attenuation is negligible. We therefore adopt the Case~B recombination value of $f(\mathrm{H}\beta) = f(\mathrm{H}\alpha)/2.86$ as the intrinsic flux estimate. Given the noise level in the NIRSpec/prism spectrum at the corresponding wavelength, this would correspond to a signal-to-noise ratio of \( \mathrm{SNR} \approx 2.4 \),  which remains consistent with the current non-detection in the NIRSpec/prism data. 
While \oiii\ is not significantly detected, we note a single-pixel excess at the expected wavelength. A forced flux measurement yields a potential flux of $(3.5 \pm 1.8) \times 10^{-19}$ erg~s$^{-1}$~cm$^{-2}$, corresponding to a $\sim$2$\sigma$ feature. 
Adopting an instrumental line width to compute the upper limit on \oiii\,$\lambda5007$, we derive a 2-$\sigma$ upper limit of $f$(\oiii$\,\lambda5007$) $< 5.6 \times 10^{-19}$ erg\,s$^{-1}$\,cm$^{-2}$, which corresponds to $R_3 < 0.89$. Using the extrapolation of the calibration from \citet{Sanders2024ApJ...962...24S}:
\begin{equation}
    \log(\mathrm{R3}) = 0.834 - 0.072x - 0.453x^2
\end{equation}
where $x = 12+\log(\rm{O/H})-8.0$, we infer a 2-$\sigma$ upper limit of $12 + \log(\mathrm{O/H}) < 6.52$ or $Z<7\times10^{-3} Z_{\odot}$. A summary of the measurements is provided in Table~\ref{tab:line}; further discussion of the metallicity can be found in Section~\ref{sec:discuss-4.1}.

\subsection{Photometry and SED Modeling}
To measure consistent total fluxes across multiple bands, we perform photometric measurements on the PSF-homogenized JWST images.
We reconstruct the NIRCam PSF utilizing the \texttt{STPSF} Python package.
To homogenize the PSF across different bands, we adopt the NIRCam F444W PSF as the target, which has the lowest spatial resolution,  and generate a convolution matching kernel using the \texttt{photutils} package, employing a 2D split cosine bell taper function to ensure smooth transitions in the kernel.
All JWST NIRCam images are then convolved with their respective kernels to standardize the PSF across all bands.
Based on the PSF-homogenized JWST images, we use the \texttt{photutils} Python package to perform aperture photometry \citep{Kron1980ApJS...43..305K} with a radius of 0\farcs15.
Then an aperture correction is applied to measure the total fluxes of \target.
For Subaru imaging in IA505, g, r, i, z, and y bands, we adopt the photometric measurements of model total flux from the COSMOS2025 catalog \citep{Shuntov2025arXiv250603243S}.
The source has a $>2$-$\sigma$ detection in IA505, g, r, i band, and non-detection in z and y band.
We also examine the deep U-band data from the CLAUDS survey \citep{Sawicki2019MNRAS.489.5202S} and B-band data from the COSMOS 20 project.
Neither the B band nor the U band presents a $>2$-$\sigma$ detection.
We adopt the 2$\sigma$ depth as an upper limit for non-detection bands.
The results of photometric measurements of \target are presented in the Table~\ref{tab:phot} and Figure~\ref{fig:img+sed}.

\begin{table}%[htpb]
\centering
\caption{Photometric Measurements}
\label{tab:phot}
\begin{tabular*}{\linewidth}{@{\extracolsep{\fill}}lcc}
\toprule
\toprule
Band & Flux density [nJy] &  AB Magnitude \\
\midrule
\midrule
&CFHT MegaCam&\\
U&$<12.1$ ($2\sigma$)&$>28.7$ ($2\sigma$)\\
\midrule
&Subaru (Hyper) Suprime-Cam&\\
SC B&$<18.4$ ($2\sigma$)&$>28.2$ ($2\sigma$)\\
SC IA505&$247.3\pm16.6$&$25.42\pm0.07$\\
HSC g&$60.8\pm3.7$&$26.94\pm0.07$\\
HSC r&$21.8\pm4.6$&$28.05\pm0.23$\\
HSC i&$13.3\pm4.6$&$28.59\pm0.37$\\
HSC z&$<31.9$ ($2\sigma$)&$>27.6$ ($2\sigma$)\\
HSC y&$<60.8$ ($2\sigma$)&$>26.9$ ($2\sigma$)\\
\midrule
&JWST NIRCam&\\
F090W&$9.9\pm4.4$&$28.90\pm0.48$\\
F115W&$14.3\pm3.9$&$28.51\pm0.30$\\
F150W&$14.3\pm1.2$&$28.51\pm0.09$\\
F200W&$13.6\pm1.9$&$28.56\pm0.15$\\
F277W&$24.9\pm1.6$&$27.91\pm0.07$\\
F356W&$4.0\pm1.4$&$29.89\pm0.37$\\
F410M&$6.3\pm3.4$&$29.40\pm0.59$\\
F444W&$14.2\pm0.8$&2$8.52\pm0.06$\\
\bottomrule
\bottomrule
\end{tabular*}
\end{table}

We conduct spectral energy distribution (SED) modeling using the \texttt{EAZY} code \citep{Brammer2008ApJ...686.1503B} and Yggdrasil Pop III models \citep{Zackrisson2011ApJ...740...13Z}.
Given that the stellar population is characterized by emission line and gas-phase metallicity analyses, we assume all stars in the galaxy are Pop III stars to estimate their stellar age and stellar mass.
Pop III templates are generated following the methodology of \citet{Fujimoto2025arXiv250111678F} to effectively sample the stellar age range and different initial mass function (IMF) scenarios.
These templates correspond to three IMF scenarios: Pop III.1 (the most top-heavy IMF), Pop III.2 (moderately top-heavy IMF), and Pop III.Kroupa (IMF of \citealt{Kroupa2001MNRAS.322..231K}).
For Pop III.1, we adopt four templates with stellar ages of $t_\mathrm{age} = [0.01, 1.0, 2.0, 3.6]$ Myr.
For Pop III.2, eight templates are used with $t_\mathrm{age} = [0.01, 1, 2, 5, 10, 19, 49, 99]$ Myr.
For Pop III.Kroupa (Kroupa IMF), eight templates are included with $t_\mathrm{age} = [0.01, 1, 2, 5, 10, 20, 50, 100]$ Myr.
In all cases, the gas covering fraction was fixed at $f_\mathrm{cov} = 1.0$ to maximize the nebular contribution.
A dust extinction of $\mathrm{E(B-V) = 0.0148}$ is used to correct the Galactic dust reddening \citep{Schlegel1998ApJ...500..525S} in the COSMOS field.
The best-fit SED model, shown in Figure~\ref{fig:img+sed}, yields a stellar age of $t_\mathrm{age} = 2~\mathrm{Myr}$ and a stellar mass of $M_\mathrm{star} = 6.1 \times 10^5~M_\odot$.
The model exhibits strong agreement with Pop III characteristics, including high EW hydrogen (H$\alpha$ traced by F277W excess) and helium emission lines (\ion{He}{1} traced by F444W excess) and a pronounced Balmer/Paschen jump (traced by faint continuum flux in F356W band).
However, we note that the broadband color F200W $-$ F150W, which traces the Balmer jump, is smoothed out by the high-EW hydrogen Balmer series emission lines.

For comparison, we perform an SED fit using only Population II models with the BAGPIPES code \citep{Carnall2018MNRAS.480.4379C}.
We employ the 2016 update \citep{Chevallard2016MNRAS.462.1415C} of the BC03 stellar population synthesis models \citep{Bruzual2003MNRAS.344.1000B} and assume a delayed-$\tau$ star formation history, a Kroupa IMF \citep{Kroupa2001MNRAS.322..231K}, and the \citet{Calzetti2000ApJ...533..682C} dust attenuation law.
Nebular emission is included with an ionization parameter in the range of $-4<\log U<-1$.
This Pop~II-only fit yields a very young ($t_\mathrm{age,2\sigma}<3.2\,\mathrm{Myr}$), low-mass ($\log M_*/M_\odot=6.8\pm0.2$), and extremely metal-poor ($Z_{2\sigma}<4.9\times10^{-3}\,Z_\odot$) galaxy with negligible dust attenuation ($A_{\mathrm{V},2\sigma}<0.03$).
Given that this model lacks Pop~III components, these inferred properties are consistent with our primary results.

\subsection{Large-scale Environment}

To investigate the large-scale environment of the \target, we measure the galaxy overdensity field using the COSMOS2025\footnote{\url{https://cosmos2025.iap.fr}}, the COSMOS-Web catalog. We use photometric redshifts derived with the \texttt{LePhare} SED code as described in the COSMOS2025 catalog paper \citep{Shuntov2025arXiv250603243S}. We restrict our overdensity measurements to sources with $m_{\mathrm{F444W}} < 26$.
For galaxies brighter than $m_{\mathrm{F444W}} < 26$ matched with spectroscopic redshifts described in the COSMOS2025 paper, the photo-$z$ performance is robust, with a median absolute deviation of $\sigma_{\mathrm{MAD}} <\sim0.02$ (defined by Eq.5 in \citealt{Shuntov2025arXiv250603243S}) and a bias of $\text{median}(\Delta z) \sim 0.005$.
At redshifts \( z = 2\text{--}4.5 \), the standard deviation of the photometric redshift uncertainties is approximately 0.1. Thus, we select galaxies within a redshift slice of \(\Delta z = \pm 0.1\) around \( z = 3.19 \), which corresponds to a comoving radial depth of approximately \(\Delta \chi \approx 183\,\mathrm{cMpc}\).

We compute the galaxy overdensity, defined as:
\begin{equation}
\delta = \frac{\rho - \bar{\rho}}{\bar{\rho}},
\end{equation}
where \(\rho\) is the local comoving number density and \(\bar{\rho}\) is the mean density averaged over all valid apertures in the redshift slice. To estimate \(\rho\), we measure the number of galaxies within a projected comoving radius of 7\,cMpc around each source in the slice, corresponding to a volume of $2.8\times10^4\,\mathrm{cMpc}^{3}$ (also see \citealt{cai2017}). In this process, we mask out regions affected by bright stars to avoid biases in the number density measurement and to mitigate photometric redshift uncertainties. We further restrict the analysis to apertures fully enclosed within the COSMOS-Web footprint. The effective comoving volume for each galaxy is calculated from the intersection of the circular aperture with the unmasked survey area.

\begin{figure}
\centering
\includegraphics[width=\linewidth]{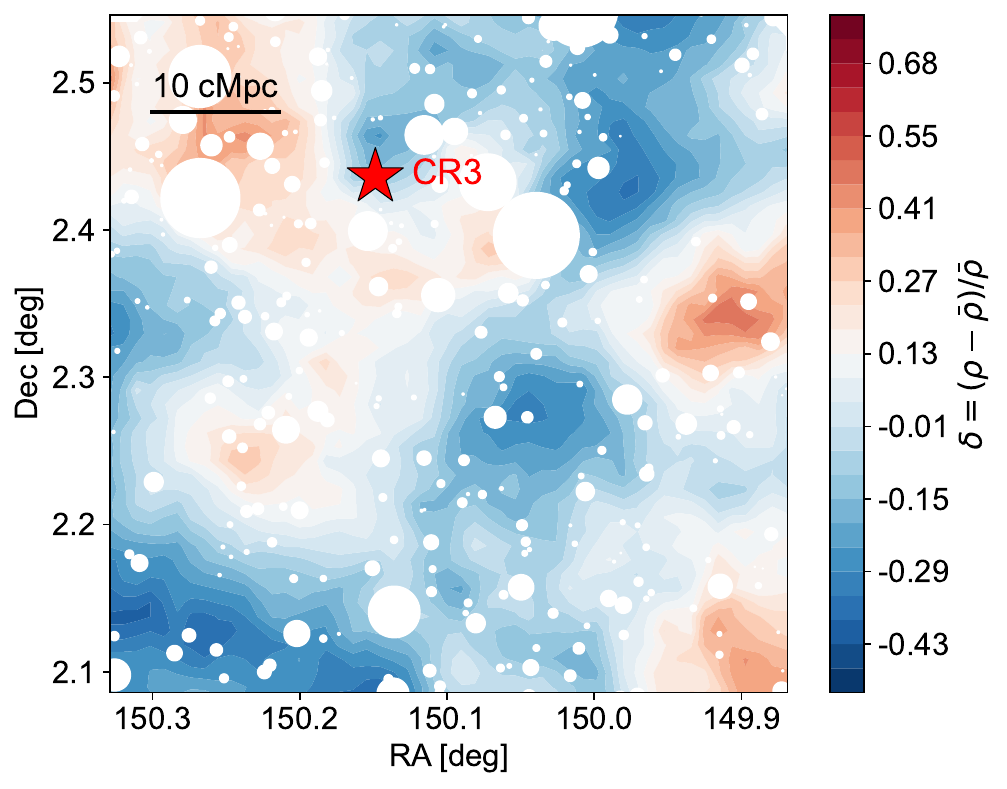} 
\caption{\textbf{Large-scale environment of the \target.}
The background shows the galaxy overdensity field at \(z = 3.2\pm0.1\) based on COSMOS2025 \texttt{LePhare} photometric redshifts, with red (blue) indicating overdense (underdense) regions. Masked regions are shown in white. 
}
\label{fig:environ}
\end{figure}

In total, we select 4608 galaxies within the redshift slice used to construct the overdensity field (see Figure \ref{fig:environ}). The standard deviation of the overdensity distribution is \(\sigma_\delta = 0.26\). The target galaxy lies at an overdensity of \(\delta = -0.12\), corresponding to approximately $-0.5$-$\sigma$ below the mean. This indicates that the \target resides in about an average or slightly underdense region of the large-scale structure.

\section{Results and Discussions} \label{sec:discuss}

\subsection{A Pop~III Candidate Growing in a Cosmic Noon Underdense Environment} \label{sec:discuss-4.1}

We identify a Pop~III galaxy candidate, the \target, at spectroscopic redshift \( z = 3.19 \). SED modeling based on Yggdrasil Pop~III templates \citep{Zackrisson2011ApJ...740...13Z} indicates a stellar age of \( t_{\mathrm{age}} = 2\,\mathrm{Myr} \) and a stellar mass of \( M_* = 6.1 \times 10^5\,M_\odot \), suggesting a young and low-mass primordial system. From rest-frame UV spectroscopy with VLT/MUSE and rest-frame optical spectroscopy with JWST/NIRSpec prism, we detect exceptionally strong emission lines of Ly\(\alpha\), H\(\alpha\), and \hei\(\,\lambda10830\), while the \heii\(\,\lambda1640\) feature coincides with an OH sky line, hindering accurate measurement. The measured fluxes and rest-frame equivalent widths are summarized in Table \ref{tab:line}.
%The measured fluxes and rest-frame equivalent widths are: \(f({\rm Ly\alpha}) = (5.8 \pm 0.7) \times 10^{-17}\,\mathrm{erg\,s^{-1}\,cm^{-2}}\), \(\mathrm{EW}_0({\rm Ly}\alpha) = 822 \pm 101\,\text{\AA}\); \(f({\rm H\alpha}) = (1.8 \pm 0.2) \times 10^{-18}\,\mathrm{erg\,s^{-1}\,cm^{-2}}\), \(\mathrm{EW}_0({\rm H}\alpha) = 2814 \pm 327\,\text{\AA}\); and $f$(\hei\(\,\lambda10830\)) \(= (4.5 \pm 1.5) \times 10^{-19}\,\mathrm{erg\,s^{-1}\,cm^{-2}}\), \(\mathrm{EW}_0(\text{\ion{He}{1}}\,\lambda18301) = 2360 \pm 807\,\text{\AA}\). 
The extremely high EW$_0$(Ly$\alpha$) is consistent with expectations for very young, metal-poor stellar populations \citep[e.g.,][]{Schaerer2003A&A...397..527S}. The strong observed H$\alpha$ is comparable to other known Pop~III candidates at \( z \sim 6 \), including LAP1 and GLIMPSE-16043 \citep{Vanzella2023A&A...678A.173V, Fujimoto2025arXiv250111678F}.

No other metal lines, such as carbon, nitrogen, or oxygen, are significantly detected, indicating an essentially unenriched, near-pristine ISM. Assuming Case B recombination, we infer an H$\beta$ flux of \(f({\rm H\beta}) = (6.3 \pm 0.7) \times 10^{-19}\,\mathrm{erg\,s^{-1}\,cm^{-2}}\). 
If we perform a forced flux measurement within a 3-pixel window centered on \hb, we obtain a flux ratio of \(f({\rm H\beta})/f({\rm H\alpha}) = 0.20\pm0.11\). This is consistent within 1.4-$\sigma$ of the Case B recombination value of $1/2.86\approx0.35$, indicating that the use of the Case B assumption to estimate the intrinsic \hb~flux is reasonable within the limitations of the current data.
The \(2\text{-}\sigma\) upper limit on the [\ion{O}{3}]$\,\lambda5007$ line flux is $f$([\ion{O}{3}]$\,\lambda5007$) \(< 5.6 \times 10^{-19}\,\mathrm{erg\,s^{-1}\,cm^{-2}} \), yielding \( R_3 < 0.89 \) and \( 12 + \log(\mathrm{O/H}) < 6.52 \), or \( Z < 7 \times 10^{-3}\,Z_{\odot} \), based on the strong-line calibration of \citet{Sanders2024ApJ...962...24S}. Although the R3 diagnostic is double-valued at low R3, the high-metallicity branch is disfavored for this very low mass galaxy.

Accounting for possible MSA flux loss, we scale the NIRSpec/prism spectrum to the F277W photometry, yielding a corrected H$\alpha$ flux of \( (4.2 \pm 0.6)\times10^{-18}\,\mathrm{erg\,s^{-1}\,cm^{-2}} \) and a Ly$\alpha$/H$\alpha$ ratio of \(13.9 \pm 2.5\). In Case B recombination, the intrinsic Ly$\alpha$/H$\alpha$ ratio is \(\sim8.7\) \citep{Brocklehurst1971MNRAS.153..471B}. The significantly elevated observed ratio suggests negligible dust attenuation, as Ly$\alpha$ photons are known to be resonantly scattered and easily suppressed in dusty media. This may be consistent with expectations for a Pop~III star formation scenario.

We acknowledge that strong-line metallicity estimates based on the R3 ratio are subject to systematic uncertainties. As emphasized by \citet{Hsiao2025arXiv250503873H}, different empirical calibrations can yield metallicities that vary by more than 0.3 dex, particularly when extrapolated to the extremely low-metallicity regime. 
For instance, the simulation-based calibration of \citet{Hirschmann2023MNRAS.526.3504H} yields a 2-$\sigma$ upper limit of 12+log(O/H) $< 6.39$.
We note that under the calibration by \citet{Nakajima2022ApJS..262....3N}, which is not optimized for high-redshift galaxies, the observed R3 value of the \target falls below the calibrated range on the low-metallicity branch, and thus does not yield a reliable metallicity estimate. While the calibration does extend to such low R3 values on the high-metallicity branch (corresponding to $Z \gtrsim Z_\odot$), we consider this solution unlikely given the low stellar mass of the galaxy.
However, using their separate calibrations for subsamples with large, medium, and small EW$_0$(H$\beta$), we infer 2-$\sigma$ upper limits of 12+log(O/H) $<$ 6.75, 6.85, and 6.95, respectively. 
Moreover, the R3 ratio is also sensitive to the ionization parameter (U); for a given R3 value, higher log(U) corresponds to an even lower inferred metallicity \citep{Nakajima2022ApJS..262....3N}. While these considerations do not weaken the conclusion that CR3 is extremely metal-poor, they highlight the need for caution in interpreting such low R3 measurements and underscore the importance of future confirmation via direct $T_e$-based metallicity diagnostics in similar systems.

We also consider the possibility that the suppressed \oiii\ and enhanced \hei$\,\lambda10830$ may result from ultra-dense gas. Similar line behavior has been observed in the ``Cliff" \citep{deGraaff2025arXiv250316600D}, a Little Red Dot (LRD) with an ultra-dense broad-line region. However, the \target shows a very different SED, and deeper, higher-resolution spectra will be needed to further investigate its gas properties.

Remarkably, this is the only known isolated galaxy at \( z \sim 3 \) with a gas-phase metallicity as low as that of the most metal-poor clumps or galaxies at \( z \gtrsim 6 \), highlighting that such low metallicity
environments can exist well after the reionization era. 
While GNHeII J1236+6215 at a similar redshift exhibits strong He\(^+\) ionization signatures, it likely traces only small pockets of young Pop~III stars embedded within otherwise evolved star-forming systems.
This finding aligns well with previous absorption-line studies of DLAs \citep[\eg,][]{Cooke2011MNRAS.412.1047C,Cooke2011MNRAS.417.1534C,Cooke2017MNRAS.467..802C,Welsh2023MNRAS.525..527W}, which have revealed extremely metal-poor gas reservoirs at similar redshifts. These suggest that pockets of pristine gas can still exist in the Universe down to redshift $z\sim3$, where conditions remain suitable for the formation of the first generation of stars.

The COSMOS2025 catalog indicates that the \target resides in an underdense large-scale environment, as inferred from photometric redshifts within a redshift slice centered at \( z = 3.19 \). 
Within a redshift slice of \(\Delta z = \pm 0.1\), consistent with the photometric redshift uncertainty and corresponding to a comoving depth of \(\sim 183\,\mathrm{cMpc}\), the \target is situated in a slightly underdense region, with an %local 
overdensity of \(\delta \approx -0.12\), approximately \(-0.5\text{-}\sigma\) below the cosmic mean.
The nearest neighboring galaxy lies at a projected proper distance of 30 kpc from the target, with a \texttt{LePhare} photometric redshift of $z = 3.26$. Given the uncertainties inherent in photometric redshifts, we currently cannot determine whether feedback from this neighbor has contributed to the chemical enrichment around CR3.

Such mildly underdense environments may help suppress external metal enrichment, allowing pockets of chemically pristine gas to persist.
Cosmological simulations have shown that metal enrichment is spatially inhomogeneous and proceeds inefficiently in low-density regions, where feedback-driven metal outflows from nearby galaxies fail to penetrate the voids \citep[\eg,][]{Tornatore2007MNRAS.382..945T, Pallottini2014MNRAS.440.2498P}. In these outskirts of the cosmic web, delayed halo assembly and suppressed star formation can allow primordial gas to survive without significant pollution, creating favorable conditions for late-time Pop~III star formation \citep[\eg,][]{Wise2012ApJ...745...50W}. Our finding may support the theoretical expectation that isolated, metal-free galaxies may be preferentially found in the underdense filaments or voids of the cosmic web, even at \( z \sim 3 \).

\subsection{Current non-detection of \ion{He}{2} Emission Lines}

For Pop~III stars, which are theoretically expected to produce extremely hard ionizing radiation fields, the detection of nebular He$^{+}$ recombination lines provides a clear and direct diagnostic of their ionizing power. Notably, \heii$\,\lambda1640$ (n=3--2, Balmer-$\alpha$) is the strongest nebular \ion{He}{2} line in the ultraviolet, while \heii$\,\lambda4686$ (n=4--3, Paschen-$\alpha$) is the strongest in the optical regime. On the zero-age main sequence, Pop~III stellar populations are expected to produce maximum rest-frame equivalent widths of $\sim100~\AA$ and $\sim150~\AA$ for \heii$\,\lambda1640$ and \heii$\,\lambda4686$, respectively \citep{Schaerer2002A&A...382...28S}.
Yet in current observations of the \target, neither line is detected, possibly for two main reasons: observation limitations and rapid stellar evolution.

First, \heii$\,\lambda1640$ lies on a strong OH sky emission line at the observed wavelength, which %severely 
hampers its recovery in the processed MUSE data. We therefore estimate a 1-$\sigma$ upper limit on its rest-frame EW of 180 \AA. Second, \heii$\,\lambda4686$ falls near $2\,\mu$m in the JWST/NIRSpec prism spectrum, where the signal-to-noise ratio is less than unity. To achieve a 3-$\sigma$ detection at this wavelength would require a rest-frame EW of $\gtrsim 570~\AA$, significantly above the expected maximum for Pop~III emission. This leaves open the possibility of detecting \heii~with space-based optical or ground-based observations offering higher sensitivity and higher spectral resolution for \heii$\,\lambda1640$. 
Further, higher-resolution near-infrared spectroscopy for \heii$\,\lambda4686$ is required in future probes.

On the other hand, another possibility is that the absence of \heii~emission results from the rapid evolution of Pop~III stars. \citet{Schaerer2003A&A...397..527S} predict that the hard ionizing radiation field produced by such stars declines quickly after star formation begins, causing the \heii~recombination lines to fade significantly after $\sim2$ Myr of continuous star formation. This timescale depends on the IMFs, and may naturally explain the low detection rate of \heii~emission lines.

\section{Conclusions} \label{sec:conclusion}

We identify a Pop~III galaxy candidate, the \target, at spectroscopic redshift \( z = 3.19 \) in the COSMOS field, based on combined JWST/NIRSpec Prism and VLT/MUSE spectroscopy, along with deep multi-wavelength imaging from JWST/NIRCam and Subaru/HSC. We summarize the key findings of this work as follows:

\begin{itemize}
\item  \textit{Strong recombination lines:} The \target exhibits exceptionally high rest-frame equivalent widths of \(\mathrm{EW}_0({\rm Ly}\alpha)=822\pm101\,\text{\AA}\), \(\mathrm{EW}_0({\rm H}\alpha) = 2814\pm327\,\text{\AA}\), and EW$_0$(\hei$\,\lambda10830$) $=2360\pm807\,\AA$, comparable to known Pop~III candidates at \(z \sim 6\) like LAP1 and GLIMPSE-16043. 

\item  \textit{No dust attenuation:} The observed Ly$\alpha$/H$\alpha$ ratio of $13.9\pm2.5$ exceeds the theoretical case~B value of 8.7, suggesting negligible dust attenuation, which may be compatible with expectations for an extremely metal-poor or metal-free environment.

\item  \textit{Lack of metal lines:} No detectable metal lines are currently found in the rest-frame UV to optical spectrum.
A marginal ($\sim$2-$\sigma$) excess is seen at [\ion{O}{3}]$\,\lambda$5007. We derive a 2-$\sigma$ upper limit of $f$(\oiii$\,\lambda$5007) \(<5.6\times10^{-19}\,\mathrm{erg\,s^{-1}\,cm^{-2}} \), yielding 
\(12 + \log(\mathrm{O/H}) < 6.52\), or \(Z < 7\times10^{-3}\,Z_\odot\), based on the \citet{Sanders2024ApJ...962...24S} calibration. Accounting for systematic uncertainties of $\gtrsim0.3$ dex in the calibrations, the most conservative 2-$\sigma$ upper limit is constrained to $12+\log(\mathrm{O/H}) < 6.95$, based on the strong-line calibration from \citet{Nakajima2022ApJS..262....3N}.

\item \textit{Extremely young stellar population:} Pop~III SED modeling indicates a stellar age of \(2\,\mathrm{Myr}\) and stellar mass of \(6.1 \times 10^5\,M_\odot\), consistent with a %very 
young, low-mass, metal-free burst of star formation.

\item \textit{Slightly underdense environment:} The \target resides in a locally average or slightly underdense region with overdensity \(\delta \approx -0.12\), about \(0.5\text{-}\sigma\) below the cosmic mean, 
suggesting potentially limited interaction with nearby sources of metal enrichment. 
\end{itemize}

This discovery implies that chemically pristine gas can persist to as late as \( z \sim 3 \), well beyond the epoch of reionization. It highlights the possibility of detecting first-generation star formation in isolated, faint galaxies during the cosmic noon. 
We have noticed a few other extremely metal-poor candidate galaxies in the COSMOS field at $z=$ 3--5 from $\sim2700$ NIRSpec spectra in the CAPERS program.
Follow-up observations are warranted. Deep space-based imaging and higher-resolution spectroscopy could enable precise measurements of metallicity and constrain helium recombination emission lines, shedding light on the pristine nature and ionizing conditions in this system.    

\begin{acknowledgments}
This work is supported by National Key
R\&D Program of China (grant no. 2023YFA1605600)
and Tsinghua University Initiative Scientific Research Program.
This work is based in part on observations made with the NASA/ESA/CSA James Webb Space Telescope. The data were obtained from the Mikulski Archive for Space Telescopes at the Space Telescope Science Institute, which is operated by the Association of Universities for Research in Astronomy, Inc., under NASA contract NAS 5-03127 for JWST. These observations are associated with program \#1727, \#1837, \#5893, \#6368, and \#6585.
The authors acknowledge the COSMOS-Web team led by CoPIs (J. Kartaltepe and C. Casey), PRIMER team led by PI (J. Dunlop), COSMOS-3D team led by CoPIs (K. Kakiichi, X. Fan, F. Wang, E. Egami, J. Lyu, and J. Yang), the CAPERS team led by PI (M. Dickinson), and the High-z Menagerie team led by CoPIs (D. Coulter, J. Pierel, and M. Engesser) for developing their observing program with a zero-exclusive-access period.
Mark Dickinson acknowledges support from program number GO-6368, provided through a grant from the STScI under NASA contract NAS5-03127.
All the JWST raw data used in this paper can be found in MAST: \dataset[10.17909/qs9h-ah96]{http://dx.doi.org/10.17909/qs9h-ah96}.
The Hyper Suprime-Cam (HSC) collaboration includes the astronomical communities of Japan and Taiwan, and Princeton University. The HSC instrumentation and software were developed by the National Astronomical Observatory of Japan (NAOJ), the Kavli Institute for the Physics and Mathematics of the Universe (Kavli IPMU), the University of Tokyo, the High Energy Accelerator Research Organization (KEK), the Academia Sinica Institute for Astronomy and Astrophysics in Taiwan (ASIAA), and Princeton University. Funding was contributed by the FIRST program from the Japanese Cabinet Office, the Ministry of Education, Culture, Sports, Science and Technology (MEXT), the Japan Society for the Promotion of Science (JSPS), Japan Science and Technology Agency (JST), the Toray Science Foundation, NAOJ, Kavli IPMU, KEK, ASIAA, and Princeton University. 
This paper makes use of software developed for Vera C. Rubin Observatory. We thank the Rubin Observatory for making their code available as free software at \url{http://pipelines.lsst.io/}.
This paper is based on data collected at the Subaru Telescope and retrieved from the HSC data archive system, which is operated by the Subaru Telescope and Astronomy Data Center (ADC) at NAOJ. Data analysis was in part carried out with the cooperation of Center for Computational Astrophysics (CfCA), NAOJ. We are honored and grateful for the opportunity of observing the Universe from Maunakea, which has the cultural, historical and natural significance in Hawaii. 
Based on observations collected at the European Southern Observatory under ESO programme 0102.A-0222(A) and data obtained from the ESO Science Archive Facility \citep{https://doi.org/10.18727/archive/41}.
This work is partly based on the COSMOS dataset \citep{https://doi.org/10.26131/irsa178}.
\end{acknowledgments}

% \begin{contribution}
% %%This section gives authors the space to recognize author contributions. The text inside this environment is NOT counted towards the total word quanta. At a minimum, manuscripts are expected to include this text:

% All authors contributed equally to the Terra Mater collaboration.

% %% But authors are expected to provide more specific details, e.g. 
% %%
% %%SC was responsible for writing and submitting the manuscript.
% %%WWM came up with the initial research concept and edited the manuscript.
% %%OTS obtained the funding and edited the manuscript.
% %%EBF provided the formal analysis and validation. He also edited the manuscript.
% %%GEH Supervised the undergraduates, wrote the software and administers the project github and Zenodo repositories.
% %%
% %% Authors can use the Contributor Role Taxonomy (CRediT) at
% %% https://credit.niso.org
% %% for ideas on how write a good statement tailored to their needs.

% \end{contribution}

%% To help institutions obtain information on the effectiveness of their 
%% telescopes the AAS Journals has created a group of keywords for telescope 
%% facilities.
%
%% Following the acknowledgments section, use the following syntax and the
%% \facility{} or \facilities{} macros to list the keywords of facilities used 
%% in the research for the paper.  Each keyword is check against the master 
%% list during copy editing.  Individual instruments can be provided in 
%% parentheses, after the keyword, but they are not verified.
\facilities{JWST (NIRCam, NIRSpec), VLT (MUSE), Subaru (Suprime-Cam, HSC)}

%% Similar to \facility{}, there is the optional \software command to allow 
%% authors a place to specify which programs were used during the creation of 
%% the manuscript. Authors should list each code and include either a
%% citation or url to the code inside ()s when available.
% \software{astropy \citep{2013A&A...558A..33A,2018AJ....156..123A,2022ApJ...935..167A},  
%           Cloudy \citep{2013RMxAA..49..137F}, 
%           Source Extractor \citep{1996A&AS..117..393B}
%           }

%% Appendix material should be preceded with a single \appendix command.
%% There should be a \section command for each appendix. Mark appendix
%% subsections with the same markup you use in the main body of the paper.
%%
%% Each Appendix (indicated with \section) will be lettered A, B, C, etc.
%% The equation counter will reset when it encounters the \appendix
%% command and will number appendix equations (A1), (A2), etc. The
%% Figure and Table counter will not reset.

%\appendix

%\section{app}\label{sec:app}

%% For this sample we use BibTeX plus aasjournalv7.bst to generate the
%% the bibliography. The sample7.bib file was populated from ADS. To
%% get the citations to show in the compiled file do the following:
%%
%% pdflatex sample7.tex
%% bibtext sample7
%% pdflatex sample7.tex
%% pdflatex sample7.tex

\bibliography{main}{}
\bibliographystyle{aasjournalv7}

%% This command is needed to show the entire author+affiliation list when
%% the collaboration and author truncation commands are used.  It has to
%% go at the end of the manuscript.
%\allauthors

%% Include this line if you are using the \added, \replaced, \deleted
%% commands to see a summary list of all changes at the end of the article.
%\listofchanges

\end{document}